\documentclass[acmsmall,screen,authorversion]{acmart}

\usepackage{subfigure}
\usepackage{cleveref}
\AtBeginDocument{%
  \providecommand\BibTeX{{%
    \normalfont B\kern-0.5em{\scshape i\kern-0.25em b}\kern-0.8em\TeX}}}
\usepackage[normalem]{ulem}

\newcommand{\revise}[1]{\textcolor{blue}{#1}}

\newcommand{\startpara}[1]{{\vskip1pt\noindent{\bf #1.}}}
\newcommand{\sectref}[1]{Section~\ref{#1}}
\newcommand{\figref}[1]{Figure~\ref{#1}}
\newcommand{\tabref}[1]{Table~\ref{#1}}

\setcopyright{acmcopyright}
\copyrightyear{2018}
\acmYear{2018}
\acmDOI{10.1145/1122445.1122456}

\acmConference[ACM Trans. Cyber-Phys. Sys]{}{}{}
\acmBooktitle{}
\acmPrice{15.00}
\acmISBN{978-1-4503-XXXX-X/18/06}

\begin{document}

\title{Planning for Automated Vehicles with Human Trust}


\author{Shili Sheng}
\affiliation{School of Engineering,  University of Virginia \institution{University of Virginia }}
\email{ss7dr@virginia.edu}
\author{Erfan Pakdamanian}
\affiliation{School of Engineering,  University of Virginia}
\email{ep2ca@virginia.edu}

\author{Kyungtae Han}
\affiliation{InfoTech Labs, Toyota Motor North America}
\email{kyungtae.han@toyota.com}

\author{Ziran Wang}
\affiliation{InfoTech Labs, Toyota Motor North America}
\email{ziran.wang@toyota.com}

\author{John Lenneman}
\affiliation{Collaborative Safety Research Center, Toyota Motor North America}
\email{john.lenneman@toyota.com}

\author{David Parker}
\affiliation{School of Computer Science, University of Birmingham}
\email{d.a.parker@cs.bham.ac.uk}

\author{Lu Feng}
\affiliation{School of Engineering, University of Virginia}
\email{lu.feng@virginia.edu}

\renewcommand{\shortauthors}{S. Sheng et al.}

\begin{abstract}
Recent work has considered personalized route planning based on user profiles, but none of it accounts for human trust. We argue that human trust is an important factor to consider when planning routes for automated vehicles. This paper presents a trust-based route planning approach for automated vehicles. We formalize the human-vehicle interaction as a partially observable Markov decision process (POMDP) and model trust as a partially observable state variable of the POMDP, representing the human's hidden mental state. We build data-driven models of human trust dynamics and takeover decisions, which are incorporated in the POMDP framework, using data collected from an online user study with 100 participants on the Amazon Mechanical Turk platform. We compute optimal routes for automated vehicles by solving optimal policies in the POMDP planning, and evaluate the resulting routes via human subject experiments with 22 participants on a driving simulator. The experimental results show that participants taking the trust-based route generally reported more positive responses in the after-driving survey than those taking the baseline (trust-free) route. In addition, we analyze the trade-offs between multiple planning objectives (e.g., trust, distance, energy consumption) via multi-objective optimization of the POMDP. We also identify a set of open issues and implications for real-world deployment of the proposed approach in automated vehicles. 
\end{abstract}



\maketitle

\section{Introduction}\label{sec:intro}
Recent years have witnessed significant advances in the development of automated vehicles, which have already been tested over millions of miles on public roads~\cite{carTest}.
However, fully autonomous vehicles that do not require human intervention are still decades away due to technology, infrastructure, and regulation limitations~\cite{carSAE5}.
The majority of  automated vehicles available to the general public nowadays are Level 2 and Level 3 of automation~\cite{sae2018taxonomy}, which allow the driver to turn attention away from the primary task of driving, but the driver must still be prepared to take over control of the vehicle when necessary. 
The human's decision on whether or not to rely on the automation is guided by trust.
Prior studies have found that distrust is the main barrier to the adoption of automated vehicles~\cite{kaur2018trust}; in addition, users with lower trust levels take over control of the vehicle more frequently~\cite{korber2018introduction}. 
On the other hand, overtrust in automation can lead to catastrophic outcomes (e.g., fatal Tesla autopilot crashes~\cite{Tesla}).
Thus, it is important to take into account the influence of human trust when developing automated vehicles. 
In this paper, we consider the problem of route planning for automated vehicles that account for trust. 

Existing route planning methods (e.g., \cite{gonzalez2007adaptive, kanoulas2006finding, andersen2013ecotour}) mostly focus on computing routes that optimize distance, time, and energy consumption metrics.
Several recent works (e.g., \cite{campigotto2016personalized,dai2015personalized,zhu2017fineroute}) consider personalized route recommendations based on user profiles (e.g., mobility options, frequently visited places).
However, none of these existing methods explicitly account for human trust. 
We argue that human trust is an important factor to consider when planning routes for automated vehicles. 
For example, if the driver has lower trust in the automated vehicle's capability for safely navigating urban streets with pedestrians constantly crossing as opposed to freeways, the driver may prefer a freeway despite longer distance.

In this work, we follow the notion of trust in automation defined in \cite{lee2004trust}, which views human trust as a delegation of responsibility for actions to the automation and willingness to accept risk (possible harm), while the decision to delegate is based on a subjective evaluation of the automation's capability for a particular task. To concretize the problem, we consider a motivating example where the automated vehicle may encounter three types of typical road incidents (i.e., pedestrian, obstacle, and oncoming truck).  
Trust is therefore affected by the human's takeover
decisions and the vehicle’s capability of handling an incident. 
We adopt the commonly used method of measuring the subjective belief of trust via user questionnaires.
Specifically, we designed and conducted an online user study with 100 participants on the Amazon Mechanical Turk platform. We asked users to watch various driving videos recorded in the driver's view
and answer questions about their trust in the automated vehicle's capability of safely handling the incident shown in the video on a 7-point Likert scale. They were also asked whether they would like to take over control of the automated vehicle, imagining that they were the driver.
We model the evolution of trust dynamics (i.e., how trust changes over time) as a linear Gaussian system using data collected from the online user study. 
We also build data-driven models to predict human takeover decisions. 

We formalize the human-vehicle interaction as a partially observable Markov decision process (POMDP),
which is a general modeling framework for planning under uncertainty~\cite{kaelbling1998planning}. 
We model trust as a partially observable state variable of the POMDP, representing the human's hidden mental state.
In addition, there are three observable state variables representing the vehicle position, the incident type, and the success/failure of the vehicle handling an incident. 
The estimated trust dynamics model informs the probabilistic transition function of the trust variable in the POMDP. 
There are two actions: the human's takeover decision and the vehicle's route choice. 
Since the vehicle does not know about the human's actual takeover decisions in advance, it assumes that humans follow the data-driven takeover decision models estimated using the online user study data. 
The goal of POMDP planning is to compute an optimal policy that makes route choices which maximize the expectation of the cumulative reward, with a reward function designed to promote better user satisfaction and safety of automated vehicles. 

We applied the proposed trust-based route planning approach to the motivating example and obtained two routes: a trust-based route where a human makes takeover decisions based on trust dynamics and incidents, and a trust-free route (as a baseline for comparison) where the human's takeover decisions only depend on incidents.
We evaluated and compared the performance of these two routes via human subject experiments on a driving simulator. 
We conducted experiments with 22 participants, who were randomly assigned to two equal-sized groups for the between-subject study (each group has 11 participants, who took one of the two routes).
The experimental results show that participants taking the trust-based route generally reported more positive responses in the after-driving survey than those taking the trust-free route.

\startpara{Contributions}
We summarize the major contributions of this work as follows.
\begin{itemize}
    \item We developed a trust-based route planning approach for automated vehicles, which is based on a POMDP framework and uses data-driven models of trust dynamics and takeover decisions.
    \item We designed and conducted an online user study with 100 participants on the Amazon Mechanical Turk platform to collect data about users' trust in automated driving.
    \item We designed and conducted human subject experiments with 22 participants on a driving simulator to evaluate the proposed approach, which showed encouraging results.
\end{itemize}
\noindent
This paper is an extended version of our previous work~\cite{sheng2021trust}. 
We add the following two new contributions.
\begin{itemize}
    \item We analyzed the trade-offs between multiple planning objectives (e.g., trust, distance, energy consumption) via multi-objective optimization of a POMDP.
    \item We discussed the limitations of the proposed approach and identified a set of open issues and implications for real-world deployment in automated vehicles.
\end{itemize}

\startpara{Paper organization}
The rest of the paper is organized as follows.
We discuss the related work in \sectref{sec:related},
describe the motivating example in \sectref{sec:motiv},
present the trust-based route planning approach in \sectref{sec:approach},
describe the driving simulator experiments in \sectref{sec:study},
analyze the multi-objective optimization results in \sectref{sec:mo},
discuss the limitations and open issues in \sectref{sec:discussion},
and draw conclusions in \sectref{sec:conclusion}.

\section{Related Work}\label{sec:related}

In this section, we survey the related work on two topics: (1) route planning for vehicles; and (2) trust in automation. For each topic, we identify gaps in the state-of-the-art and discuss the connection with this paper. 

\subsection{Route Planning for Vehicles}
The goal of route planning is to compute the optimal routes for vehicles. 
The most commonly used metrics include distance, travel time, and fuel consumption. 
Graph search algorithms such as Dijkstra's algorithm~\cite{dijkstra1959note} and  $A$* algorithm~\cite{hart1968formal} can be applied to find the shortest distance path between any two locations. 
Computing the fastest route (i.e., with the least travel time) is more challenging than finding the shortest distance route.
\citet{kanoulas2006finding} extended the $A$* algorithm by considering the speed change at a different time of the day to compute the fastest route.
\citet{gonzalez2007adaptive} developed an adaptive fastest route planning method based on information learned from historical traffic data, accounting for various factors (e.g., road quality, weather condition, area crime rate) that may influence vehicle speed patterns.
\citet{andersen2013ecotour} proposed to find the most eco-friendly route by assigning eco-weights based on GPS and fuel consumption data. 

There are several recent studies considering personalized route recommendation for users. 
\citet{campigotto2016personalized} developed a method for personalized route planning by using Bayesian learning to update users' profiles such as home location, workplace, and mobility options.
\citet{dai2015personalized} recommended a personalized optimal route considering user preferences encoded as a ratio between different metrics such as distance, travel time, and fuel consumption.
\citet{zhu2017fineroute} proposed a personalized and time-sensitive route planning method, in which they inferred users' preferences with locations and visiting time through historical data.
 
None of the aforementioned route planning methods considers human trust. 
In this paper, we develop a trust-based route planning approach to fill this gap.

\subsection{Trust in Automation}
Trust in the context of human-technology relationships can be roughly classified into three categories: (1) \emph{credentials-based}, which is used mainly in security and determines if a user can be trusted based on a set of credentials~\cite{kagal2001trust};
(2) \emph{experience-based}, which includes reputation-based trust in peer-to-peer and e-commerce applications, and determines an agent's trust value based on its own experience in predicting the probability of the execution of a certain action by another agent~\cite{krukow2008trust}; and (3) \emph{cognitive trust}, which explicitly accounts for not only the human experience, but also subjective judgment about preferences and mental states~\cite{falcone2001social}. 
In this paper, we are interested in human trust in automated vehicles, and therefore consider the cognitive trust that captures the human notion of trust.
More precisely, we follow the notion of \emph{trust in automation} proposed in~\cite{lee2004trust}, which indicates a human's willingness to rely on automation.

Studies have found that human trust changes over time during the interaction with automation, affected by various factors such as the automation's reliability, predictability, and transparency~\cite{schaefer2016meta,hancock2011meta}. 
Studies have also shown that trust can influence a human's reliance on automation,  and
the system is likely to be under-utilized if humans mistrust the automation~\cite{dzindolet2003role}.
For example, a recent study found that users with lower trust tended to take over control from  automated vehicles more frequently~\cite{korber2018introduction}. 
Inspired by insights from these prior studies, we develop a data-driven trust dynamics model to represent the evolution of human trust in automated vehicles and a takeover decision model to associate the likelihood of human's takeover decision with trust. 

Different methods to measure trust have been proposed. 
User questionnaires are commonly used to evaluate the subjective belief of trust~\cite{martelaro2016tell,xu2015optimo}.
For example, the study in~\cite{choi2015investigating} asked questions about users' trust in automated vehicles on a 7-point Likert scale. 
In addition, various sensing technologies have been used for the continuous measurement of human trust in real-time, including gaze tracking~\cite{hergeth2016keep}, gestures (e.g., face touching and arms crossed)~\cite{lee2011modeling}, and biometrics (e.g., electroencephalogram and galvanic skin response)~\cite{hu2016real}.
We measure human trust on a 7-point Likert scale via questionnaires in the online user study, and via continuous user control input (i.e., pressing buttons mounted on the steering wheel) in the driving simulator study.

Existing work on trust in automated vehicles includes investigating factors that influence users' adoption of automated vehicles~\cite{lee2019assessing,lee2019exploring, sheng2019case}, studying the effect of alarm timing on drivers' trust~\cite{abe2004effect}, designing forward-collision warning system~\cite{koustanai2012simulator} and cruise control system~\cite{cahour2009does} to improve users' trust.
By contrast, this paper develops a route planning approach that accounts for trust to improve the user experience of automated vehicles.

Several recent works have explored the idea of modeling trust with POMDPs.
For example, a POMDP model for trust-workload dynamics in Level 2 driving automation was developed in~\cite{akash2020toward}, and a POMDP-based method for human-robot collaboration in table cleaning tasks was proposed in~\cite{chen2020trust}.
Our work is inspired by these methods, but differs from them in the following aspects. 
First, we focus on applying trust-based planning for automated vehicles, which requires different POMDP modeling from existing work.  
Second, we designed and conducted human subject experiments based on driving simulations for data collection and model evaluation. 
Further, we use multi-objective optimization of POMDPs to analyze the trade-offs between multiple planning objectives (e.g., trust, distance, energy consumption).

\section{Motivating Example}\label{sec:motiv}
\begin{figure}[t]
   \centering
   \includegraphics[width=0.85\linewidth]{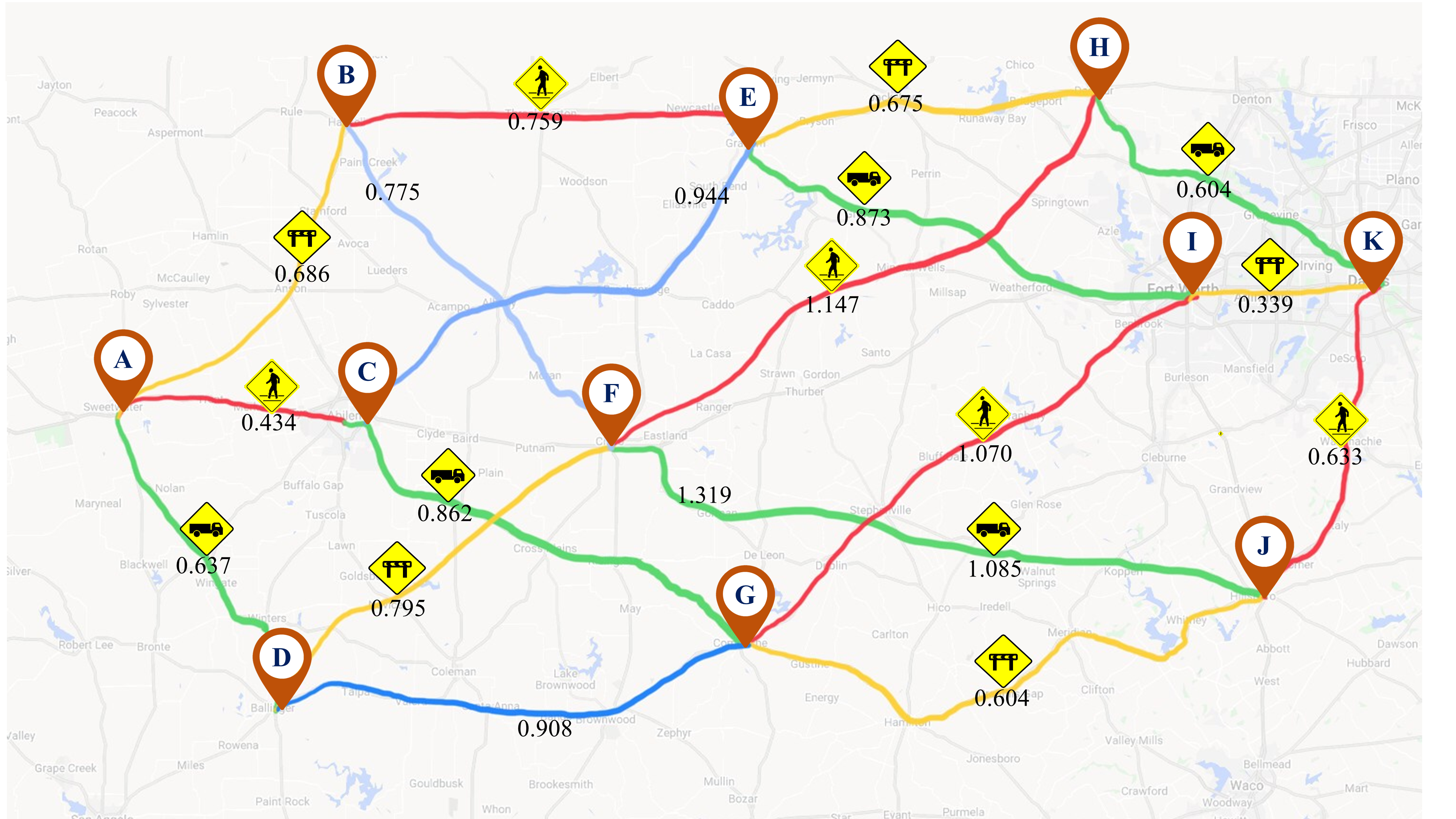}
   \caption{An example map with three types of road incidents (pedestrian, obstacle, and oncoming truck). The number on each road segment indicates its distance (1 unit = 10 miles).}
   \label{fig:map}
\end{figure}

We describe a motivating example of route planning for automated vehicles.
\figref{fig:map} shows an example map, where three types of typical incidents that may occur on the road are considered:
(1) a pedestrian crossing the road, (2) an obstacle ahead of the lane, and (3) an oncoming truck in the neighboring lane. 
We can easily generalize to more complex examples with a richer set of incidents. 
For simplicity, we assume that each road segment may have up to one incident at a time. 
We also assume that the vehicle has information about the potential incident that it may encounter in the next road segment. Such information can be easily obtained, for example, via sensing and crowdsourcing traffic monitoring apps.

\begin{figure*}[t]
   \centering
   \includegraphics[width=0.9\linewidth]{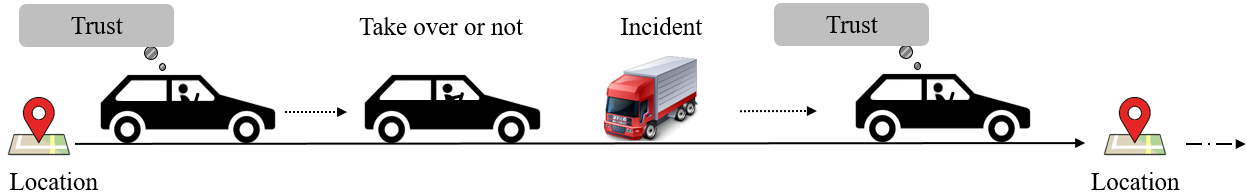}
   \caption{A schematic view of an automated vehicle navigating from one location to another. When approaching an incident, the driver may decide to take over and switch to manual driving. The takeover decision can be influenced by the driver's trust in the automated vehicle, which evolves over time.}
   \label{fig:schematic}
\end{figure*}

\figref{fig:schematic} shows a schematic view of the automated vehicle traveling from one location to another. Suppose that the vehicle is approaching an incident in autopilot mode. Due to safety concerns, the driver may decide to take over control of the vehicle and switch to manual driving. Such takeover decisions can be influenced by the driver's trust in the automated vehicle's capability of handling different types of incidents: the driver with lower trust is more likely to take over.  
In addition, the driver's trust evolves over time depending on the takeover decision and the vehicle’s capability of handling an incident.

\emph{The goal of this work is to develop a trust-based route planning approach that computes an optimal route for the automated vehicle (e.g., navigating from A to K in the example map) while taking into account human trust dynamics and takeover decisions.} 

\section{Trust-based Route Planning}\label{sec:approach}
We present a trust-based route planning approach for automated vehicles. 
The key idea is to model the human-vehicle interaction as a POMDP and compute the optimal vehicle route by solving the optimal policy using POMDP planning. 

\subsection{The Proposed POMDP Framework}\label{sec:pomdp}
Formally, a POMDP is denoted as a tuple $(S, A, \mathcal{T}, R, O, \mathcal{\delta}, \gamma)$, 
where $S$ is a finite set of states, $A$ is a set of actions, $\mathcal{T}$ is the transition function representing conditional transition probabilities between states, $R: S \times A \to \mathbb{R}$ is the real-valued reward function, $O$ is a set of observations, $\mathcal{\delta}$ is the observation function representing the conditional probabilities of observations given states and actions, and $\gamma \in [0,1]$ is the discount factor.   
At each time step $t$, given an action $a_t \in A$, a state $s_t \in S$ evolves to $s_{t+1} \in S$ with probability $\mathcal{T}(s_{t+1} | s_t, a_t)$. The agent receives a reward $R(s_t, a_t)$, and makes an observation $o_{t+1} \in O$ about the next state $s_{t+1}$ with probability $\delta(o_{t+1} | s_{t+1}, a_t)$. 
The goal of POMDP planning is to compute the optimal policy \revise{$\pi^{*}$} that chooses actions to maximize the expectation of the cumulative reward $\mathbb{E} [\sum_{t=0}^\infty \gamma^t R(s_t, a_t)]$.

\begin{figure}[b]
   \centering
   \includegraphics[width=0.3\linewidth]{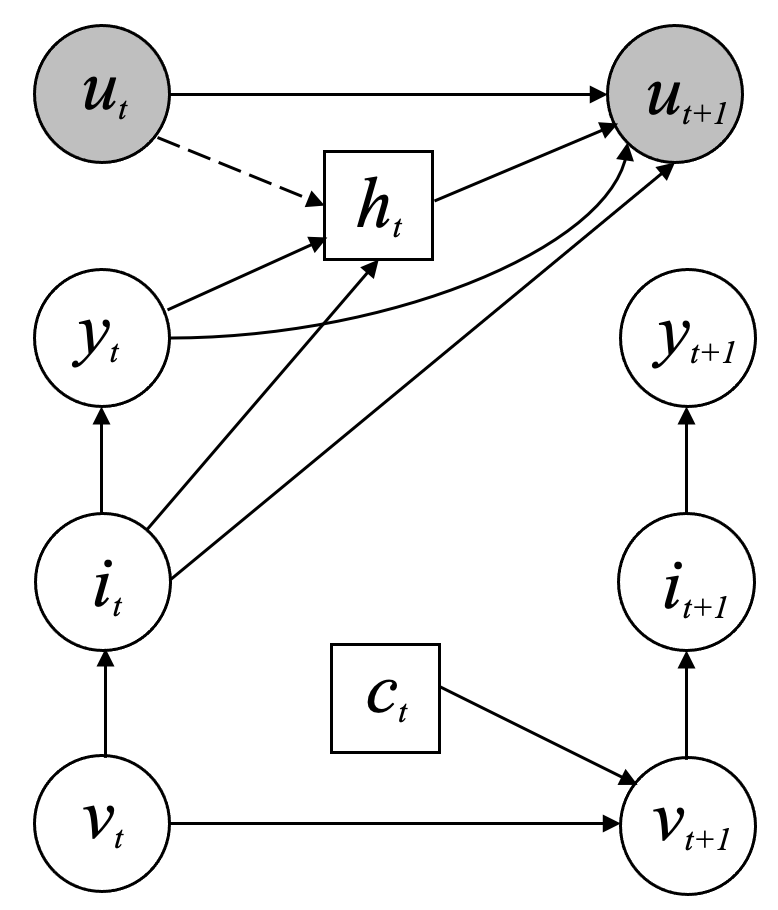}
   \caption{The POMDP graphical model for trust-based route planning. (Each node represents a state variable. Shadowed nodes are partially observable variables. Squares represent actions. Arrows represent transition functions.) }
   \label{fig:pomdp}
\end{figure}

\figref{fig:pomdp} illustrates a graphical model of the proposed POMDP framework for trust-based route planning. We factor the state $s_t$ at time $t$ into four variables: $v_t$ represents the vehicle position, $i_t$ represents the road incident, $y_t$ represents the automated vehicle's capability of safely handling the incident, and $u_t$ is a partially observable variable representing the human's trust in the automated vehicle (because trust is a hidden human mental state that cannot be directly observed by the vehicle agent).
We factor the action $a_t$ at time $t$ into two variables: the vehicle route choice $c_t$ and the human's takeover decision $h_t$.
Given the vehicle's current position $v_t$ and the route choice action $c_t$, we can determine the next vehicle position $v_{t+1}$ by the transition function $\mathcal{T}(v_{t+1}|v_t, c_t)$.
The potential incident $i_{t}$ that the vehicle may encounter is determined by the vehicle position with probability $\mathcal{T}(i_t | v_t)$, and the automated vehicle's capability of safely handling the incident $i_t$ is given by $\mathcal{T}(y_t | i_t)$.
As discussed in \sectref{sec:related}, trust in automation can be influenced by many factors. 
Here, we model the evolution of trust dynamics with a probabilistic transition function $\mathcal{T}(u_{t+1}|u_t, y_t, i_t, h_t)$, based on a simplified assumption that trust evolves depending on the takeover decision and the vehicle's capability of handling an incident. 
The intuition is that trust may increase when the human chooses to not take over and witnesses the automated vehicle successfully handling an incident, and the trust may decrease if the automated vehicle fails to handle an incident. 
We set the POMDP discount factor as $\gamma =1$.

The vehicle agent does not know about the human's actual takeover action in advance, and it computes the optimal POMDP policy $\pi^*$ of route choices $c_t$ based on a model that predicts the human's takeover decision $h_t$. 
We consider two different takeover decision models for comparison:
(1) a \emph{trust-free model}, denoted by $\pi^h(h_t | i_t, y_t)$, where the human decides whether to takeover depending on the incident and a fixed belief on the automated vehicle's capability to handle certain types of incidents;
and (2) a \emph{trust-based model}, denoted by $\pi^h(h_t | i_t, y_t, u_t)$, where a human makes takeover decisions based on the incident and trust, indicating that the human's belief in the automated vehicle's capability changes over time depending on the trust dynamics.

Consider the motivating example described in \sectref{sec:motiv}. 
The vehicle position $v_t$ is one of the locations $\{A, B, \dots, K\}$ shown in the map (\figref{fig:map}). 
The incident $i_t$ can take one of the four values: null, pedestrian, obstacle, and truck. 
The vehicle's capability $y_t$ of handling incidents has binary outcomes: success and failure. 
Since the human's trust is a partially observable variable $u_t$ representing the hidden mental state, we use an observation variable $\hat{u}_t$ to represent the subjective trust on a 7-point Likert scale (1 and 7 indicate the lowest and highest levels of trust, respectively) measured via user questionnaires. 
The available route choices $c_t$ are given by the map. 
For example, in location $A$, the vehicle may choose one of the three routes colored in yellow, red, and green to navigate to $B$, $C$, or $D$, respectively. 
The human takeover decision $h_t$ is a binary choice of whether or not to take over control of the vehicle and resume manual driving. 
We can define the transition functions $\mathcal{T}(v_{t+1}|v_t, c_t)$ and $\mathcal{T}(i_t | v_t)$ based on the map. 
We can estimate $\mathcal{T}(y_t | i_t)$ based on the historical testing logs of the automated vehicle safely handling incidents. For the motivating example, we assume that the automated vehicle can always safely handle incidents (but the human driver has no prior knowledge about this assumption).

\begin{table}[b]\centering
\caption{POMDP reward function} 
\label{tab:reward} 
\begin{tabular}{l|c|c|c c}
\toprule
         & Pedestrian & Obstacle & Truck   \\ \hline \hline
Autopilot (Success)  & 3 & 2  & 1            \\\hline
Autopilot (Failure)  & -9    & -6  & 0       \\ \hline
Manual driving & 0     & 0        & 0                   \\\bottomrule 
\end{tabular}
\end{table}

We design a reward function shown in \tabref{tab:reward} for the motivating example. 
Intuitively, we want to reward for better safety and user satisfaction of automated vehicles. 
If the automated vehicle handles an incident successfully, we assign positive rewards based on the difficulty of driving tasks. 
When approaching a pedestrian incident, the automated vehicle needs to stop before the crosswalk and wait till the pedestrian crossing the road. 
When approaching an obstacle incident, the automated vehicle needs to perform lane changing in order to avoid a collision with the obstacle. 
When there is an oncoming truck in the neighboring lane, the automated vehicle needs to keep driving in the same lane. 
Thus, we rank the pedestrian incident as the most difficult task and assign the highest reward value of 3, followed by the obstacle incident with a reward value of 2 and the truck incident with a reward value of 1.
On the other hand, if the automated vehicle fails to handle an incident safely, we assign rewards based on the severity of the incident (e.g., striking a pedestrian can cause more serious damage than colliding with an obstacle).
We assign zero reward to manual driving, because we want to promote better a user experience and let the driver enjoy non-driving tasks (e.g., reading or using mobile devices) in the automated vehicle. 
In addition, we assign a reward value of 5 to an empty road (i.e., no incident, thus no failure or takeover) to indicate that this is the most favorable choice.

For the rest of this section, we describe the design of an online user study for data collection in \sectref{sec:amt};
we present a data-driven method to estimate trust dynamics $\mathcal{T}(u_{t+1}|u_t, y_t, i_t, h_t)$ 
and the observation function $\delta (\hat{u}_t | u_t)$ in \sectref{sec:trust};
we describe the data-driven modeling of trust-free takeover decision $\pi^h(h_t | i_t, y_t)$ and the trust-based takeover decision $\pi^h(h_t | i_t, y_t, u_t)$ in \sectref{sec:takeover};
and finally, we apply the proposed approach to the motivating example and present the computed optimal routes in \sectref{sec:example}.

\begin{center}
    \begin{figure*}[t]
    \centering
      \includegraphics[width=0.98\textwidth,scale=1.0]{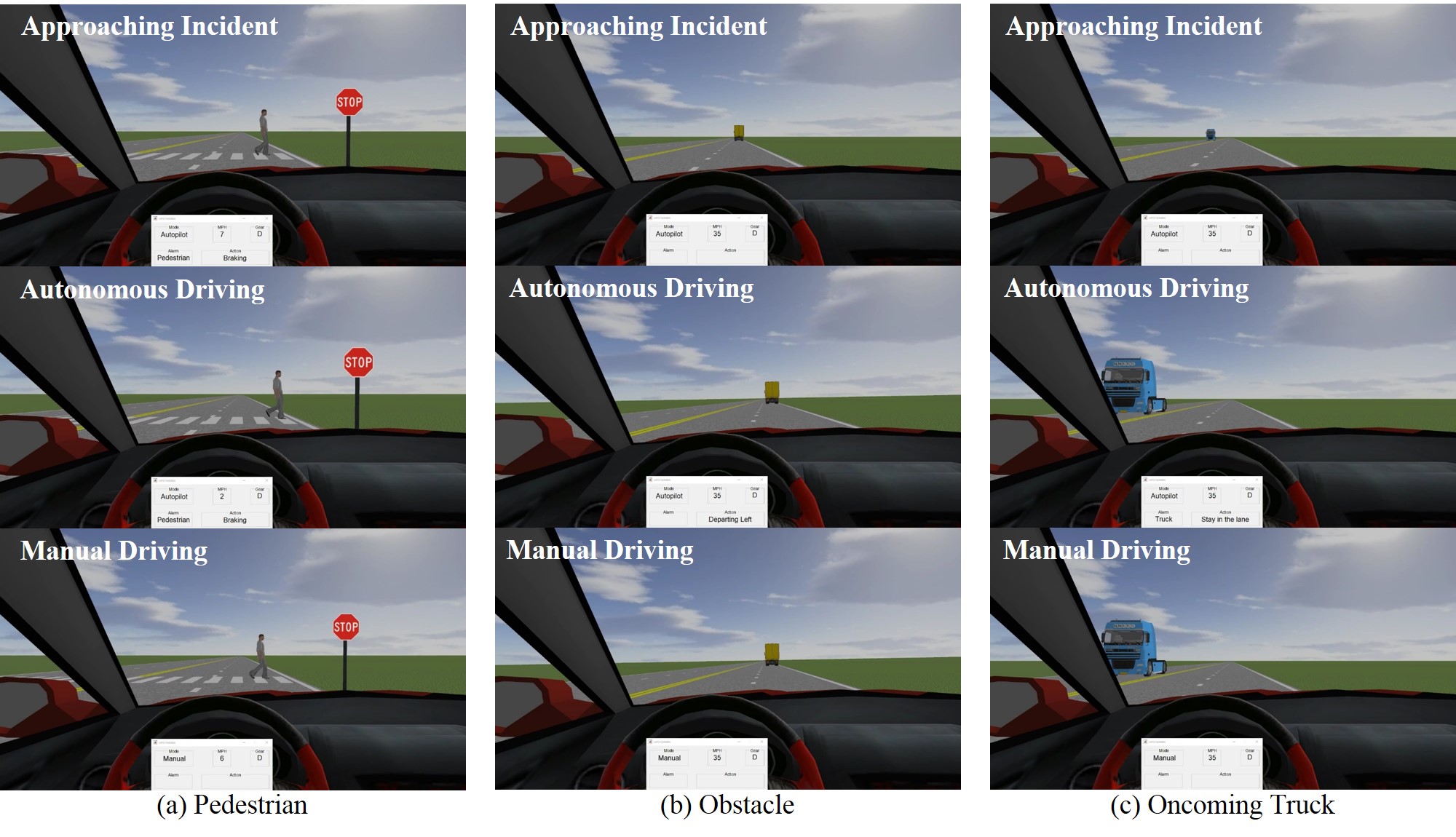}
      \caption{Screenshots of driving videos used in the online user study, covering three types of incidents: (a) a pedestrian crossing the road, (b) an obstacle (a stopped truck) ahead of the lane, (c) an oncoming truck in the neighboring lane. Each sub-figure shows: (top) the driver's view when the automated vehicle is approaching the incident, (middle) the view of autonomous driving if the driver chooses not to take over, (bottom) the view of manual driving if the driver chooses to take over.}
      \label{fig:AMT}
    \end{figure*}
\end{center}

\subsection{Online User Study for Data Collection}\label{sec:amt}
We designed and conducted an online user study\footnote{This study was approved by the Institutional Review Board (IRB) at the University of Virginia.} with 100 anonymous participants on the Amazon Mechanical Turk platform. 
The objective of this study is to collect data about human trust in automated vehicles. 
In particular, we investigated how trust evolves with respect to different incidents on the road and how a human's takeover decisions are affected by incidents and trust. 
We created a set of driving videos using the PreScan driving simulation software~\cite{PreScan}. 
\figref{fig:AMT} shows screenshots of example videos covering three types of incidents (i.e., pedestrian, obstacle, and oncoming truck) used in the motivating example.

During the online user study, we first established the baseline by asking participants about their trust in automated vehicles on a 7-point Likert scale (i.e., trust ranges from 1 to 7). 
Then, we showed a video of the automated vehicle approaching an incident on the road from the driver's view,
and asked participants if they would like a takeover control of the vehicle and switch to manual driving, imagining that they were the driver sitting inside the automated vehicle. 
Depending on the participant's response, we showed the next video of the vehicle being driven either autonomously or manually to handle the incident.
After that, we asked participants to fill in a questionnaire which estimates their updated trust in the automated vehicle. We adapted Muir's questionnaire~\cite{muir2002operators} and asked participants to answer the following questions on 7-point Likert scale:
\begin{enumerate}
    \item To what extent can you predict the automated vehicle's behavior from moment to moment?
    \item To what extent can you count on the automated vehicle to do its job? 
    \item What degree of faith do you have that the automated vehicle will be able to cope with similar incidents in the future?  
    \item Overall, how much do you trust the automated vehicle?
\end{enumerate}
We averaged a participant's responses to these four questions into a single rating between 1 and 7 to represent the participant's updated trust. 
We repeated the above process nine times (three times per incident type) with  randomized order of incidents.

We did not include any vehicle crash or near-crash videos in this study due to IRB restrictions on the ethical obligation and potential risks (e.g., some participants may feel uncomfortable watching such videos).
However, participants were not aware of such information in advance.
Instead, we instructed them to make takeover decisions based on their trust beliefs about the automated vehicle's capability to safely handle certain incidents, which may vary based on their prior experience.

The data we collected from each participant has the following format:
$\{\hat{u}_0,i_0,h_0,\hat{u}_1,\dots,i_8,h_8,\hat{u}_9\}$,
where $\hat{u}_t$ is the measured user trust, $i_t$ is the incident type, $h_t$ is the user decision of takeover or not, at each time step $t$.
Our study recruitment criteria required that participants must be able to read English fluently and have an above 95\% approval rate on the Amazon Mechanical Turk platform. 
We also inserted questions for attention checks during the study to guarantee the data quality.

\subsection{Data-Driven Trust Dynamics Model}\label{sec:trust}
As described in \sectref{sec:pomdp}, the proposed POMDP framework for trust-based route planning represents human trust as a partially observable variable $u_t$ at time step $t$, which evolves to $u_{t+1}$ over time depending on the human's takeover decision $h_t$ and the automated vehicle's capability $y_t$ to handle incident $i_t$.
Using the data collected from the online user study described in \sectref{sec:amt}, 
we model the trust dynamics and the POMDP observation function as a linear Gaussian system:
$$\mathcal{T}(u_{t+1}|u_t, y_t, i_t, h_t)=\mathcal{N}(\alpha_t u_t+\beta_t, \sigma_t^2)$$
$$\hat{u}_t \sim \mathcal{N}(u_t, \sigma_u^2)$$
where $\mathcal{N}(\mu,\sigma^2)$ represents the Gaussian distribution with the mean $\mu$ and the variance $\sigma$;
$\alpha_t$ and $\beta_t$ are linear coefficients of trust dynamics given $y_t$, $i_t$ and $h_t$; 
and $\hat{u}_t$ represents the observations of trust measured via subjective questionnaires in the online user study.
We estimate these parameter values using full Bayesian inference with Hamiltonian Monte Carlo sampling algorithm~\cite{duane1987hybrid}.

\begin{center}
    \begin{figure}[t]
    \centering
      \includegraphics[width=0.85\linewidth]{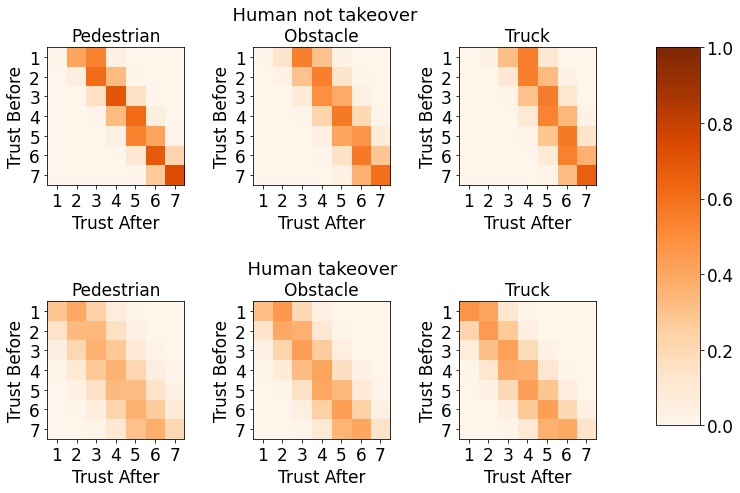}
      \caption{Visualization of probabilistic transition matrices of the learned trust dynamics model, where $u_t$ and $u_{t+1}$ are shown as trust before and trust after values ranging from 1 to 7, and each matrix corresponds to a pair of incident and takeover decision.}
      \label{fig:trust}
    \end{figure}
\end{center}

\figref{fig:trust} illustrates a visualization of the learned trust dynamics model.
There are six probabilistic transition matrices, corresponding to all combinations of three road incidents and binary human takeover decisions. 
Each transition matrix indicates the probability of changing from $u_t$ (trust before value) to $u_{t+1}$ (trust after value).
We observe that trust values are more likely to increase when a human decides not to take over (top row of \figref{fig:trust}),
while trust values tend to be constant or decrease when there is a takeover decision (bottom row of \figref{fig:trust}).
These observations are consistent with the insight from the prior studies (see \sectref{sec:related}) that takeover decisions are often correlated to trust. 
 
\subsection{Data-Driven Takeover Decision Models}\label{sec:takeover}
In the POMDP framework, we use the variable $h_t$ to denote the human's takeover decisions (i.e., whether or not to take over control of the vehicle) when approaching an incident $i_t$ at time step $t$. 
Such takeover decisions may also be influenced by human trust $u_t$. 
In the following, we present two takeover decision models based on whether or not to consider trust as an influencing factor.

\startpara{Trust-free takeover decision model}
Let $b^i$ denote human's belief in the automated vehicle's capability of safely handling an incident $i$, which remains constant in the trust-free model. 
Let $p_t$ denote the probability of the human deciding not to take over at time step $t$.
We define $p_t=\mathcal{S}(b^i r^{\mathsf{s},i} + (1-b^i) r^{\mathsf{f},i})$,
where $\mathcal{S}(x)= \frac{1}{1+e^{-x}}$ is the sigmoid function, 
$r^{\mathsf{s},i}$ and $r^{\mathsf{f},i}$ are rewards of the automated vehicle handling the incident $i$ with success and failure (see \tabref{tab:reward}), respectively. 
We model the takeover decision with a Bernoulli distribution, denoted by $h_t \sim \mathcal{B}(p_t)$.

\startpara{Trust-based takeover decision model}
Let $b^i_t$ denote the human's belief in the automated vehicle's capability of safely handling an incident $i$ at time step $t$, which evolves over time depending on the human trust $u_t$.
Thus, we model the belief as a sigmoid function $b^i_t=\mathcal{S}(\kappa^i u_t + \lambda^i)$, 
where $\kappa^i$ and $\lambda^i$ are linear coefficients associated with the incident $i$.
We assume that the human trust $u_t$ follows a Gaussian distribution,
denoted by $\hat{u}_t \sim \mathcal{N}(u_t, \sigma_u^2)$ where $\hat{u}_t$ are the measured trust values from the online user study.
We define the probability of the human deciding not to takeover as 
$p_t=\mathcal{S}(b^i_t r^{\mathsf{s},i} + (1-b^i_t) r^{\mathsf{f},i})$, which is defined similarly to the trust-free model, but using the dynamic belief $b^i_t$ instead of the constant $b^i$.
Finally, the takeover decision is given by the Bernoulli distribution $h_t \sim \mathcal{B}(p_t)$.

\startpara{Data-driven modeling results}
We applied full Bayesian inference with the Hamiltonian Monte Carlo (HMC) sampling algorithm~\cite{duane1987hybrid} to estimate parameters in both the trust-free and trust-based models, using the data collected from the online user study. 
In Bayesian inference, Markov chain Monte Carlo (MCMC) methods are often used to obtain samples from a probability distribution.  HMC improves the previous MCMC methods on random walk by simulating a physical system using Hamiltonian dynamics.  We refer to~\cite{betancourt2017conceptual} for further introduction of HMC. We employed Stan statistical computation software~\cite{carpenter2017stan} for the implementation of the Bayesian inference through HMC.

\begin{figure}[t]
    \centering
   	\includegraphics[width=0.7\linewidth]{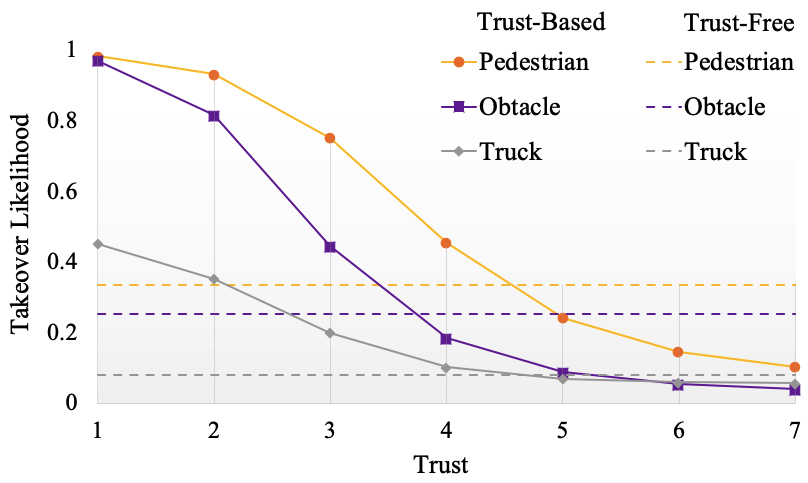}
   	\caption{Predictions of takeover likelihood with respect to trust and incidents, using trust-based and trust-free takeover decision models.}
   	\label{fig:takeover}
\end{figure}

The results of log-likelihood show that the trust-based model (-359.37) fits the collected data better than the trust-free model (-446.83). The difference in log-likelihood results shows that accounting for trust in the takeover decision model can achieve better prediction performance, which supports our assumption that human takeover decisions are influenced by trust.
\figref{fig:takeover} shows model predictions of takeover probability with respect to trust and incidents. With the trust-free model, since the takeover decision does not depend on human trust, we observe three straight lines for three incidents. With the trust-based model, we observe the general trends of decreasing takeover likelihood with increasing trust, which is consistent with findings in the prior studies (see \sectref{sec:related}).  
Furthermore, we observe from the results of both models that it is more likely for a human to decide to take over with riskier incidents: pedestrian with the highest takeover probability, followed by obstacle and truck.

\subsection{Planning for the Motivating Example}\label{sec:example}
We applied the Approximate POMDP Planning (APPL) Toolkit~\cite{SARSOP}, which is an implementation of the point-based SARSOP algorithm for efficient POMDP planning~\cite{kurniawati2008sarsop}, to compute the optimal policies of the proposed POMDP framework. 
For the motivating example, depending on the use of trust-based and trust-free takeover decision models, we obtained two optimal routes:
\begin{itemize}
    \item trust-based route: A-D-G-J-K
    \item trust-free route: A-C-E-H-K
\end{itemize}
Note that the main difference between these two routes is the order of road incidents.
In the trust-based route, the ordered incidents occurring in each road segment are: oncoming truck (A-D), null (D-G), obstacle (G-J), and pedestrian ~(J-K).
In the trust-free route, the incidents follow the order of: pedestrian (A-C), null (C-E), obstacle (E-H), and oncoming truck  ~(H-K).  
We evaluate and compare the performance of these two routes via human subject experiments\footnote{This study was approved by the Institutional Review Board at the University of Virginia.} on a driving simulator, as described in the next section.

\section{Driving Simulator Experiments}\label{sec:study}
We describe the design, procedure, and results of our driving simulator experiments as follows. 

\begin{figure}[b]
   	\centering
   	\includegraphics[width=0.75\linewidth]{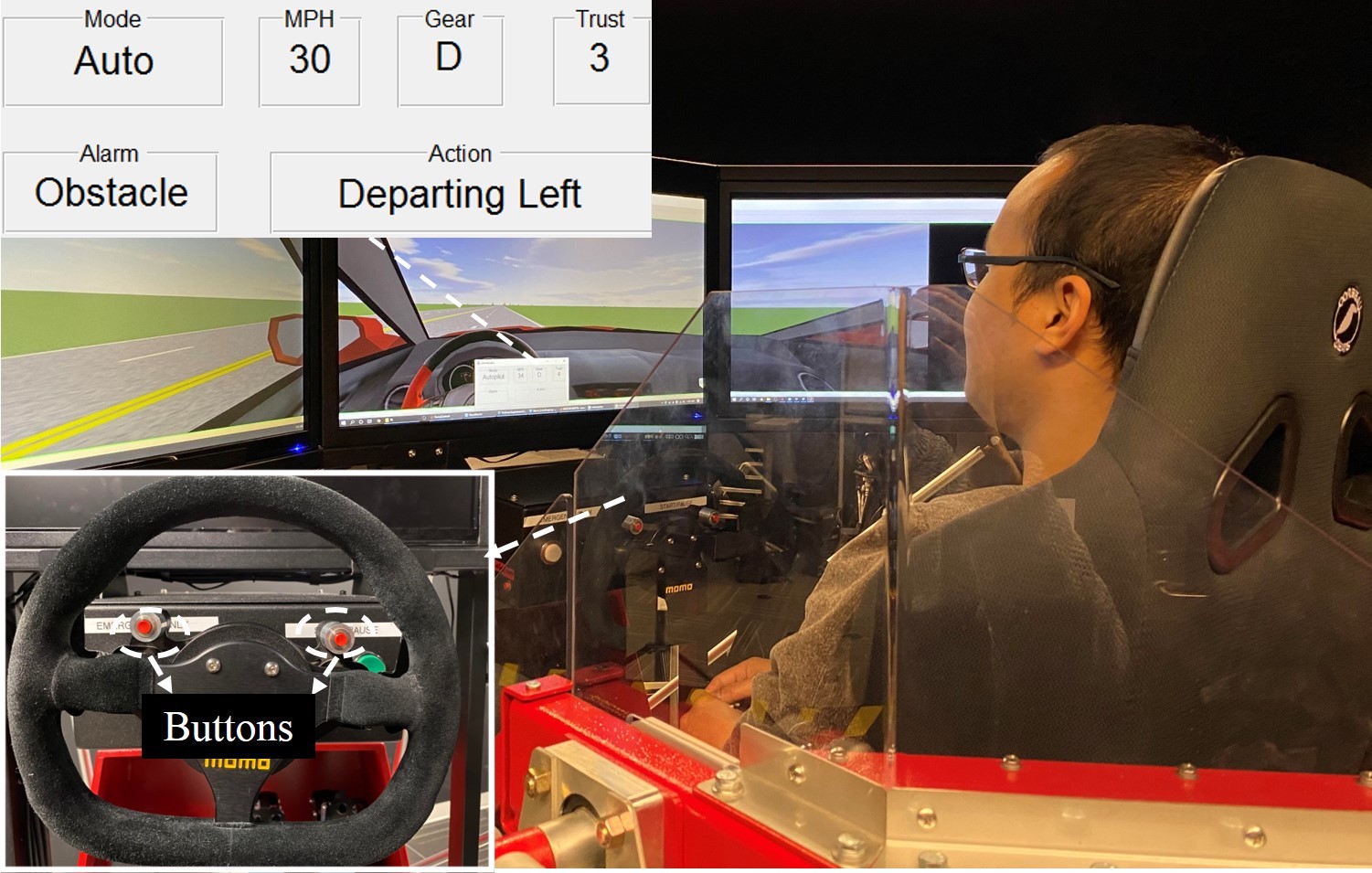}
   	\caption{Driving simulator setup. The top zoomed-in view shows the GUI displaying the driver's current trust value, along with other information such as driving mode, speed, gear, incident alarm, vehicle action. The bottom zoomed-in view shows the steering wheel with buttons for takeover commands and user trust input. }
   	\label{fig:simulator}
\end{figure} 

\subsection{Experiment Design}

\startpara{Apparatus}
\figref{fig:simulator} shows the driving simulator setup used for the experiments. 
The hardware platform is based on the Force Dynamics 401CR driving simulator, which is a four-axis motion platform that tilts and rotates to simulate the experience of being in a vehicle. The platform includes the seat, interlocked seat belt, interlocked doors, display screen, steering wheel, brake, paddle shifters, and throttle.
There are two buttons mounted on the steering wheel (bottom zoomed-in view in \figref{fig:simulator}). 
We programmed the simulator's control input such that the driver can switch between automated and manual driving by pressing the two buttons simultaneously. 
In addition, we used the same set of buttons to measure participants' trust in automated vehicles during the experiments. 
The driver can press the left (\emph{resp.} right) button to decrease (\emph{resp.} increase) the trust value ranging from 1 to 7.
\startpara{Driving scenario}
We created a driving scenario based on the motivating example described in \sectref{sec:motiv}, using the PreScan driving simulation software~\cite{PreScan}. 
We also programmed an autopilot controller for the simulated automated vehicle, which has the capability of leveraging the integrated sensors (e.g., radar, Lidar, and GPS) in PreScan for various driving tasks such as lane keeping, detecting and handling incidents.

\startpara{Manipulated factor}
We manipulate a single factor: the route that the autopilot controller follows. As stated in \sectref{sec:example}, the two conditions are: trust-based route and trust-free route.

\startpara{Dependent measures}
We are interested in studying the route which brings more cumulative reward. We recorded the participants' takeover decisions and calculated the cumulative POMDP reward using the reward function defined in \tabref{tab:reward}.

\startpara{Hypothesis}
We hypothesize that participants taking the trust-based route can obtain higher cumulative POMDP rewards than those taking the trust-free route. 

\startpara{Subject allocation}
We recruited 22 participants (average age: 23.7 years, SD=4.3 years, 31.8\% female) from the university community. 
Each participant was compensated with a \$20 gift card for completing the experiment. 
The recruitment criteria required all participants to have a valid driver's license, at least one year of driving experience, and regular or corrected-to-normal vision.
To avoid participants' bias, we adopted a between-subject study design: we randomly allocated 11 participants to take the trust-based route and the other 11 participants to take the trust-free route.

\subsection{Experiment Procedure}
Upon arrival, a participant was instructed to read and sign a consent form approved by the Institutional Review Board. 
We conducted a five-minute training session to familiarize the participant with the driving simulator setup. 
Then, the participant was instructed to drive through the trust-based or trust-free route with the simulated automated vehicle, depending on the assigned study group. 
The journey started in autopilot mode. 
When the vehicle approached an incident (i.e., pedestrian, obstacle, or truck), it alerted the participant by issuing an auditory alarm and displaying textual information about the incident type in the GUI. 
If the participant decided not to takeover, the vehicle would continue in the autopilot mode to handle the incident.
The participant can take over control of the vehicle and switch to manual driving at any point during the experiment. 
If the participant did takeover, he was required to switch back to autopilot mode after the vehicle passing that incident.
We asked the participant to periodically record their trust in the automated vehicle using the buttons on the steering wheel (see bottom left in \figref{fig:simulator}). 
After the driving session, we asked the participant to answer the following survey questions on a 7-point Likert scale (1 means strongly disagree, four is neural, seven means strongly agree).
\begin{itemize}
    \item[Q1] I believe that the automated vehicle can get me to the destination safely.
    \item[Q2] I find the route easy to drive.
    \item[Q3] I find it easy to take over control of the automated vehicle.
    \item[Q4] I have a concern about using the automated vehicle to drive through this route.
    \item[Q5] I believe that the selected route is not dangerous.
    \item[Q6] I think the selected route fits well with the way I would like to drive.
    \item[Q7] I can depend on the reliability of the automated vehicle.
\end{itemize}
It took about 40 minutes for each participant to complete the entire experiment.

\subsection{Results} 
\begin{figure}[t]
   \centering
   \includegraphics[width=0.4\linewidth]{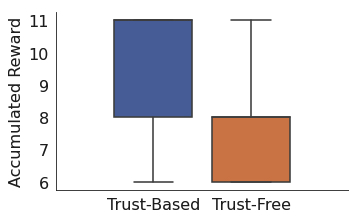}
   \caption{The cumulative rewards of participants taking trust-based and trust-free routes. }
   \label{fig:exp-reward}
\end{figure}

We calculated the cumulative POMDP rewards (using the reward function defined in \tabref{tab:reward}) for each participant, based on their takeover decisions when approaching incidents along the route.
\figref{fig:exp-reward} shows the box plot of cumulative rewards for all participants. 
We observe that participants taking the trust-based route tend to achieve higher cumulative rewards than participants taking the trust-free route, which is consistent with our study hypothesis. 
We also performed one-way analysis of variance (ANOVA) to evaluate this hypothesis, i.e, comparing the observed $F$-test statistics with $F(d_1,d_2)$ ($F$-distribution with  between-group degree of freedom $d_1$ and within-group degree of freedom $d_2$). The observed statistics $F(1,20)=9.14$ is greater than the critical value at significance level 0.01.
Thus, our study hypothesis is supported by ANOVA results statistically.

\begin{figure}[t]
   	\centering
   	\includegraphics[width=0.45\linewidth]{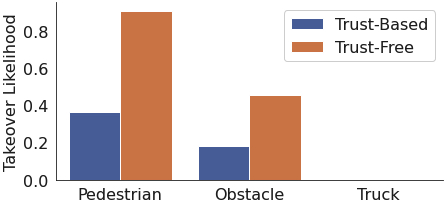}
   	\caption{Participants' average takeover likelihood when the vehicle approaching different incidents in the trust-based and trust-free routes.  }
   	\label{fig:exp-takeover}
\end{figure}

\figref{fig:exp-takeover} shows the average takeover likelihood of all participants for different incidents along the two routes. 
It is not surprising to find that participants are more likely to take over in the trust-free route than the trust-based route.
With both routes, participants have higher probabilities to take over when approaching a pedestrian than an obstacle, while none of them choose to take over the control when there was an oncoming truck in the neighboring lane.  
A possible explanation is that participants are more likely to take over when approaching incidents that are more challenging to handle or can cause more severe damages. 
These trends are consistent with the takeover predictions computed using the online user study data (see \figref{fig:takeover}).

\begin{figure}[t]
  	\centering
  	\includegraphics[width=0.6\linewidth]{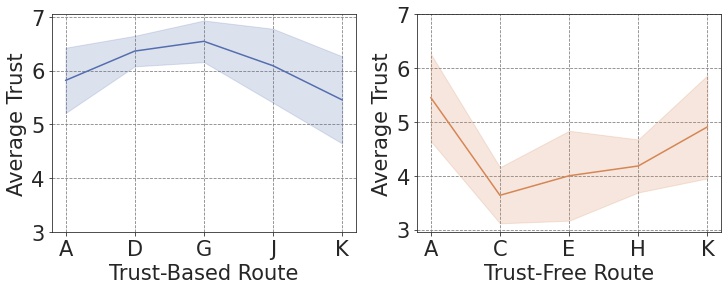}
  	\caption{The evolution of participants' average trust along the trust-based and trust-free route. (The shadow represents the 95\% confidence interval.)}
  	\label{fig:exp-trust}
\end{figure}

\figref{fig:exp-trust} shows how participants' average trust in the automated vehicle evolves as they drive through different locations along the two routes. 
For the trust-based route, we observe that the average trust increases in the route segment A-D; this may result from the automated vehicle successfully handling the incident of the oncoming truck in this segment. 
The trust continues to increase in the segment D-G, which is an empty road without any incident. 
However, the trust decreases in the next segment G-J where the vehicle needs to change lanes to avoid an obstacle, and the trust further decreases in the last segment J-K where the vehicle needs to stop and wait for a pedestrian to cross the road. 
The decrease in average trust may be explained by the occurrence of more challenging and riskier incidents.
For the trust-free route, we observe that the average trust drops sharply in the first route segment A-C with a pedestrian incident. However, the trust continues to increase slowly for the rest of the route. 
The average trust of participants taking the trust-based route is generally higher than taking the trust-free route. 

\begin{figure}[t]
   \centering
   \includegraphics[width=0.7\linewidth]{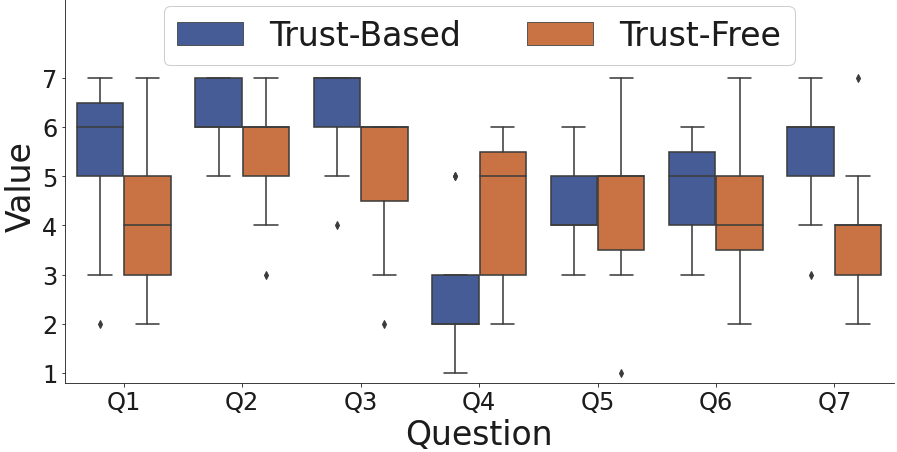}
   \caption{After-driving survey results. (Each box plot shows the maximum, the first quartile, the median, the third quartile, and the minimum. Each dot represents an outlier.)}
   \label{fig:exp-survey}
\end{figure}

\figref{fig:exp-survey} summarizes the participants' responses to the after-driving survey questions. 
The results of Q1 indicate that participants who experienced the trust-based route had higher belief in the automated vehicle's capability of driving safely than participants who experienced the trust-free route. 
The results of Q2 show that participants found the trust-based route easier to drive than the trust-free route.
The results of Q3 illustrate that participants driving through the trust-based route found it easier to take over control of the vehicle than those driving through the trust-free route. 
The results of Q4 show that participants who experienced the trust-based route had less concern about the automated vehicle than those who experienced the trust-free route.
The results of Q5 indicate that participants tended to have a neutral opinion about how dangerous the routes are.
The results of Q6 show that participants thought the trust-based route fits the way they would like to drive better than the trust-free route in general. 
The results of Q7 find that participants driving through the trust-based route perceived higher reliability of the automated vehicle than those who experienced the trust-free route. 

In summary, our human subject experimental results show that 
\begin{itemize}
    \item Participants taking the trust-based route generally resulted in higher cumulative POMDP rewards (where the reward function was designed to promote better user satisfaction and safety of automated vehicles) than those taking the trust-free route.
    \item Participants were more likely to take over in the trust-free route than in the trust-based route, and riskier incidents led to higher takeover likelihood.
    \item Participants' trust in the automated vehicle evolved during the driving experience and was influenced by different types of incidents. 
    \item Participants experienced the trust-based route had more positive responses in the after-driving survey than those driving through the trust-free route. 
\end{itemize}

\section{Multi-Objective Optimization Analysis}\label{sec:mo}

In the previous sections, we computed optimal POMDP policies based on a reward function (\tabref{tab:reward}) designed to promote better user satisfaction and safety of automated vehicles. 
In reality, users may want to achieve multiple planning objectives at once (e.g., minimizing the distance while maximizing user satisfaction) when it comes to choosing the best route for automated vehicles.
Thus, in this section, we apply multi-objective optimization to the proposed POMDP framework and analyze the trade-offs between various planning objectives.

\subsection{Objectives for Route Planning}\label{sec:objectives}

We consider the following typical route planning objectives in this section.
Each objective is modeled as a different reward function, described below~\footnote{See the concrete model file at \url{https://www.prismmodelchecker.org/files/tcps-trust}},
and the expected cumulative value of the rewards are either minimized or maximized.

\begin{itemize}
    \item \emph{User satisfaction}: defined by the reward function described in \sectref{sec:pomdp} and shown in \tabref{tab:reward}. 
    \item \emph{Distance}: modeled with a reward function that gives the distance of each route segment (as annotated in \figref{fig:map}), determined by the route choice $c_t$ made in each vehicle position $v_t$.
    \item \emph{Energy consumption rate}: we compute the average rate, by summing the energy consumption per unit distance for each route segment and dividing the sum by the total distance; energy consumption depends on the human's takeover decision, where we assume that the energy consumption rate of manual driving is 1.25 times higher than automated driving, since the latter is likely to be more energy efficient.
    \item \emph{Total energy consumption}: modeled with a reward function that gives the energy usage for each route segment; this is computed as the distance multiplied by the energy consumption rate, with the latter as described above.
    \item \emph{Average trust}: computed by summing the driver's trust level $u_t$ at each waypoint and dividing the sum by the total number of waypoints along the entire route.
    \item \emph{Trust at destination}: computed via a one-off reward attached to the final route segment, whose value is the driver's trust level $u_n$ and $n$ is the time step of reaching the destination; since trust evolves dynamically along the route, this value is likely to be different from the average trust, both of which would influence the driver's adoption of automated vehicles in the future.
\end{itemize}

\subsection{Pareto Optimal Solutions}\label{sec:pareto}

Multi-objective optimization seeks to balance the trade-offs between multiple objectives, where a single global solution that optimizes each individual objective simultaneously may not exist.
If user preferences about the relative importance of objectives are known (represented as weights over objectives), a multi-objective optimization problem can be reduced to a single-objective optimization problem by taking the weighted sum of those objective values~\cite{marler2004survey,roijers2014bounded}.
When user preferences are not specified \emph{a priori} (sometimes it is difficult to come up with weights over objectives), a set of \emph{Pareto optimal} solutions (i.e., those for which no objective can be optimized further without worsening some other objective) can be computed to assist decision-making.

We implemented a prototype procedure to compute Pareto optimal solutions for POMDPs based on the PRISM model checker~\cite{KNP11}. Our implementation is based on PRISM's POMDP solver~\cite{NPZ17} and a sampling of objective weights. Note that we switch POMDP solvers, because the APPL tool used in \sectref{sec:example} does not directly support multi-objective optimization of POMDPs.

\begin{figure}[t]
   \centering
   \includegraphics[width=0.95\linewidth]{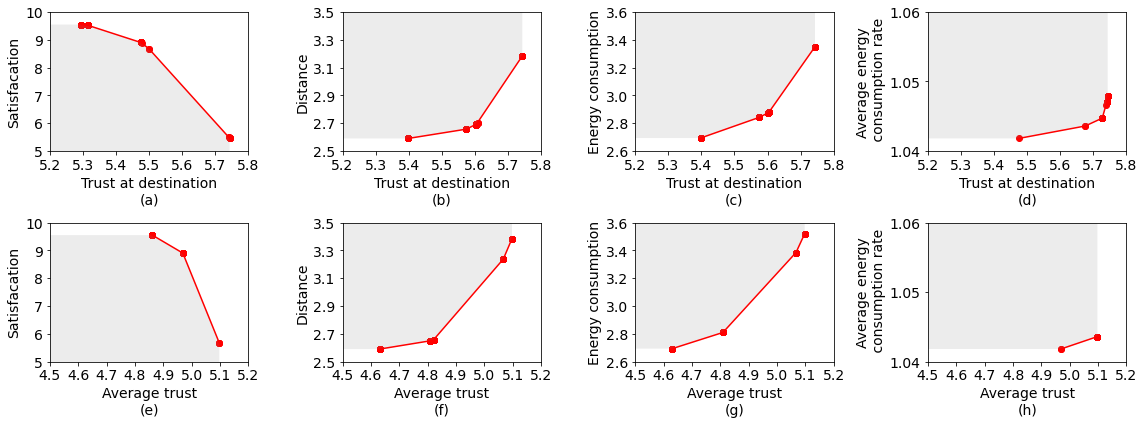}
    \caption{Pareto optimal solutions of multi-objective POMDP planning. The red line represents the Pareto curve. Each dot on the red line represent a Pareto optimal policy of the POMDP. Any point in the gray area represents a pair of objective values that can be achieved by a feasible POMDP policy. }
   \label{fig:pareto_2}
\end{figure}

We apply the prototype implementation to the multi-objective planning of objectives described in \sectref{sec:objectives}. 
\figref{fig:pareto_2} plots the Pareto optimal solutions of multi-objective POMDP planning considering various combinations of planning objectives. 
Specifically, the top row of \figref{fig:pareto_2} illustrates the trade-offs between maximizing the trust at destination vs. (a) maximizing the user satisfaction, (b) minimizing the distance, (c) minimizing the total energy consumption, and (d) minimizing the energy consumption rate;
the bottom row of \figref{fig:pareto_2} shows the trade-offs between maximizing the average trust vs. (e) maximizing the user satisfaction, (f) minimizing the distance, (g) minimizing the total energy consumption, and (h) minimizing the energy consumption rate.
As shown in \figref{fig:pareto_2}, there does not exist a global solution that optimizes each pair of objectives at the same time. Instead, users may be presented with these Pareto optimal solutions to choose a point (on the Pareto curve) that corresponds to a Pareto optimal policy for the POMDP.

Different user preferences over objectives can result in different POMDP policies (i.e., routes for vehicles). 
Let the weight vector $\vec{w}=(w_1,w_2)$ denote the user preference over two different objectives, where $0 \le w_1, w_2 \le 1$ and $w_1 + w_2=1$. 
We choose a sample  of weight vectors $\vec{w}$ by iteratively considering all values of $w_1$ in the range $[0, 1]$ in increments of 0.01, and taking $w_2=1-w_1$.
We then compute the optimal policy $\pi^{*}_{\vec{w}}$ for each $\vec{w}$ by solving the POMDP with the reward function $R_{\vec{w}} =w_1R_1+w_2R_2 $, where $R_1$ and $R_2$ are the reward functions for each objective. 
  We compute a corresponding value vector $\vec{V}=[V_1, V_2]$, where $V_1$ and $V_2$ denote the expected values of each objective for the policy $\pi^{*}_{\vec{w}}$.
  We thus obtain a set of value vectors for the set of weight vectors,
  which consititute the points on the Pareto curve.

Suppose that the planning objectives are to (1) minimize the distance, while (2) maximizing the trust at the destination, as shown in \figref{fig:pareto_2}(b). When $\vec{w} = (1,0)$, the resulting Pareto optimal point is the bottom left red dot shown in \figref{fig:pareto_2}(b), which yields a POMDP policy corresponding to the shortest route A-C-E-I-K in the map shown in \figref{fig:map}. 
When $\vec{w} = (0,1)$, the resulting Pareto optimal point is the top right red dot in \figref{fig:pareto_2}(b), which yields a different route A-D-F-H-K with longer distance but higher trust at the destination. 
When $\vec{w} = (0.5, 0.5)$, the resulting Pareto optimal point is the middle red dot in \figref{fig:pareto_2}(b), which yields the route A-B-E-I-K seeking to balance the trade-offs between two objectives. 

\section{Discussion}\label{sec:discussion}

\subsection{Limitations}
There are several limitations of this work.
First, we only consider three types of typical road incidents (i.e., pedestrian, obstacle, and oncoming truck). While it is straightforward to extend the POMDP framework with a richer set of incidents, we will need to design and conduct new online user studies to collect data about trust in the automated vehicle's capability of safely handling these new incident types and build new data-driven trust dynamics model. 

Second, we only apply our approach to the motivating example with a small map featuring a limited number of waypoints. We believe that the proposed POMDP framework can be applied to larger problems (e.g., larger maps, more locations, and more route choices). For example, the SARSOP algorithm used in \sectref{sec:example} is able to scale up the POMDP problems with $10^5$ states. However, the bottleneck lies in the evaluation. We will need to design and conduct new human subject experiments to evaluate the resulting routes of these new problems, which can be costly and time-consuming.

\subsection{Open Issues and Implications on Real-World Deployment}
We envision that this work can contribute to route planning in future automated vehicles, which would account for human trust dynamics and the trade-offs between multiple planning objectives (e.g., distance, energy consumption, safety, user satisfaction). 
However, the following open issues need to be addressed before making the proposed approach ready for real-world deployment.

The first issue is: how do we measure, calibrate and model individual drivers' trust dynamics in real time? 
In this work, we build a data-driven trust dynamic model based on the aggregated data collected from 100 participants of an online user study. In a real-world deployment, each individual's trust dynamics may vary for different drivers and change over time. There is a need for building personalized trust dynamic models and calibrating the model using real-time sensing data about human trust. 
In addition, there are challenges such as lowering the barrier of entry (e.g., using low-cost hardware and software) for collecting real-time human sensing data in vehicles, and how to guarantee the privacy of collected human data and its usage in model learning and planning.

The second issue is: how do we compute POMDP policies for large-scale planning problems in complex, adaptive traffic conditions that automated vehicles may encounter in the real world?
These challenges would require not only improving the scalability of POMDP solvers, but also the computational efficiency in order to obtain planning results in real time. 
One promising direction is to consider online POMDP algorithms (e.g., DESPOT~\cite{somani2013despot}) that have been successfully implemented for real-time autonomous driving~\cite{bai2015intention}. 
However, there is a lack of online POMDP algorithms for multi-objective optimization.

\section{Conclusion}\label{sec:conclusion}
In this paper, we present a trust-based route planning approach for automated vehicles.
We model the human-vehicle interaction as a POMDP and compute optimal routes for the vehicle by solving the POMDP planning problem. 
In order to incorporate trust into route planning, we build data-driven models of trust dynamics and takeover decisions using data collected from an online user study with 100 participants on the Amazon Mechanical Turk platform.
We applied the proposed trust-based route planning approach to a motivating example and obtained a trust-based route and a trust-free route (a baseline for comparison).
We evaluated these two routes via human subject experiments with 22 participants on a driving simulator.
The results show that participants taking the trust-based route generally resulted in higher cumulative POMDP rewards (where the reward function was designed to promote better safety and user experience of automated vehicles), were less likely to take over control of the vehicle, and reported more positive responses in the after-driving survey than those taking the trust-free route. 
We also observed that participants' trust changed over time during the study and was influenced by different road incidents. These observations are consistent with the findings of prior studies.
In addition, we analyze the trade-offs between multiple planning objectives (e.g., trust, distance, energy consumption) via multi-objective optimization of POMDP. We also identify a set of open issues and implications on the real-world deployment of the proposed approach in automated vehicles.

This work makes the first step towards incorporating human trust into route planning for automated vehicles. 
There are a few directions for future work. 
First, we would like to consider a richer set of incident types to reflect the complex road conditions that automated vehicles may encounter in the real world. 
Second, we would like to improve and evaluate the scalability of the proposed approach.
Furthermore, we would like to explore the POMDP modeling of other factors that may influence human trust in automated vehicles, such as system transparency, vehicle speed, driving styles, and user's situational awareness.  
Finally, we would like to investigate personalized modeling of individual driver's trust dynamics.

\begin{acks}
This work was supported in part by National Science Foundation grants CCF-1942836 and CNS-1755784, and European Research Council (ERC) under the European Union’s Horizon 2020 research and innovation programme (grant agreement No. 834115, FUN2MODEL)").
\end{acks}

\bibliographystyle{ACM-Reference-Format}
\bibliography{references}

\end{document}